\input harvmac
\Title{\vbox{
\hbox{HUTP-98/A033}
\hbox{\tt hep-th/9804131}}}{Extending Mirror Conjecture to
Calabi-Yau with Bundles}
\bigskip
\centerline{Cumrun Vafa}
\bigskip
\centerline{Lyman Laboratory of Physics}
\centerline{Harvard University}
\centerline{Cambridge, MA 02138, USA}
\vskip .3in

We define the notion of mirror of a Calabi-Yau manifold
with a stable bundle in the context of type II strings in terms
of supersymmetric cycles on the mirror.
This allows us to relate the variation of Hodge
structure for cohomologies arising from the bundle
to the counting of holomorphic maps of Riemann
surfaces with boundary on the mirror side.  Moreover it opens up
the possibility of studying bundles on Calabi-Yau manifolds in terms of
supersymmetric
cycles on the mirror.

\Date{April 1998}
\newsec{Introduction}
Mirror symmetry, which was conjectured as a generalization
of $R\rightarrow 1/R$ duality for Calabi-Yau compactifications
\ref\lvw{W. Lerche, C. Vafa and N.P. Warner,
Nucl. Phys. {\bf B324} (1989) 427.}\ref\ld{L. Dixon, unpublished.}\  has
played a major role in  our understanding of dynamical
issues in string theory.  At the level of string
perturbation theory, the concrete examples
of mirror pairs \ref\gpl{B. Greene and M.R. Plesser, Nucl.
Phys. {\bf B338} (1990) 15.}\ was shown to result
in a deeper understanding of sigma model instantons \ref\cand{Candelas
et. al., Nucl. Phys. {\bf B359} (1991) 21.}.
Non-perturbatively mirror symmetry also plays a key role in the
geometric engineering approach to constructing quantum field theories.
The construction of mirror pairs has been generalized using toric methods
\ref\bat{V. Batyrev, J. Alg. Geom. {\bf 3} (1994) 493.}
and some proposals exist as to how
one may derive these constructions starting from $R\rightarrow 1/R$ duality
\ref\syz{A. Strominger, S.-T. Yau and E. Zaslow, Nucl. Phys. {\bf B479}
(1996) 243.}\ref\Morr{D. Morrison,
Proc. European Algebraic Geometry Conference
(Warwick, 1996), alg-geom/9608006.}\ref\gross{M. Gross
and P.M.H. Wilson, alg-geom/9608004.}\ref\lev{N.C. Leung
and C. Vafa, hep-th/9711013}.
The basic idea is to view in a certain degenerate limit the Calabi-Yau
$n$-fold as locally a fiber space with fiber an $n$ dimensional real
torus, and applying T-daulity fiber by fiber.
The purpose of this note is to extend the mirror conjecture
to situations including bundles over Calabi-Yau.  The basic
idea of this extended conjecture was motivated by the topological
open and closed strings on Calabi-Yau threefolds \ref\wcs{E. Witten,
hep-th/9207094.}\ref\bcov{Bershadsky et. al., Comm. Math. Phys. {\bf 165}
(1994) 311.
}\ and was in fact discussed in a preliminary
form
in \ref\va{C. Vafa, Talk given at IAS, Spring 1994.}.  The sharpened version
of the present conjecture is motivated by the recent advances
in our understanding of D-branes \ref\pol{J. Polchinski,
Phys. Rev. Lett. {\bf 75} (1995) 4724.}.  See also the related works
\ref\kon{M. Kontsevich, Proceedings of the 1994 ICM {\bf I},
Birk\"auser, Z\"urich, 1995, p. 120; alg-geom/9411018.}\ref\fuka{
K. Fukaya, The Proceedings of the 1993 GARC Workshop on
Geometry and Topology, H.J. Kim, ed., Seoul National University.}\ref\zas{
A. Polishchuk and E. Zaslow, math.AG/9801119.}.

Bundles over Calabi-Yau 3-folds have phenomenological applications
for constructing $N=1$, $d=4 $ vacua for heterotic string. In this
context they lead to $(2,0)$ sigma model on the worldsheet and
one could attempt to construct a mirror in this context.  Some
proposals have been made in this setup \ref\kd{J. Distler and S.
Kachru, Nucl. Phys. {\bf B442} (1995) 64.}\ref\sha{E. Sharpe,
hep-th/9804066.}.
This is not what we will be considering in this paper.  We will
concentrate on bundles in the context of wrapped D-branes over Calabi-Yau
manifolds and discuss their mirror.  Some of the constructions are most
natural for the Calabi-Yau threefold which will also be the main
focus of this paper, though generalizations are straight-forward
for any $n$-fold Calabi-Yau and some of our general discussion
is in that context.

\newsec{Basic Review of Standard Mirror Symmetry}
We start with a Calabi-Yau 3-fold $M$ and its mirror pair
$W$.  Very roughly speaking $M$ and $W$ are fiber spaces
with fiber $T^3$ and the main difference between them is that
in one case the radii of $T^3$ are inverted.  This is related
to proposal of \syz\ for viewing the mirror Calabi-Yau as
 the moduli of a supersymmetric $T^3$ cycle.
This view of the mirror pair has been developed in \lev\ with
the result that one can give an intuitive physical explanation
of constructing mirror pairs using dual toric data \bat .

One of the basic consequences of mirror symmetry is that curve
counting becomes very simple.  Let $t^i$ parameterize
the Kahler moduli of $M$ relative to a basis
of 2-cycles $[C_i]$, i.e. $t_i=\int_{C_i}k$ where
$k$ is the Kahler form.   Moreover let
$q_i=e^{-t_i}$. Let $d^{\{ n_l \} }$ denote
the number of holomorphic maps from the sphere to Calabi-Yau,
whose image is in $\sum n_i[ C_i]$.  On the other hand
consider $W$ and the unique (up to overall scale)  holomorphic 3-form on it
$\Omega$. Then the basic mirror conjecture in particular implies
that
\eqn\gnm{\sum n_in_j n_k d^{\{ n_l \} }{\prod_l q_l^{n_l}\over (1-\prod_l
q_l^{n_l})}=
\int_W \partial_i \partial_j\partial_k \Omega \wedge \Omega}
for suitable choice of coordinates on moduli of complex structure
of $W$ (known as special coordinates) and normalization of $\Omega$ .
The denominator in the left-hand side of the
above formula comes from contribution
of multi-wrapped instantons in $M$.

\newsec{Extension of Mirror Conjecture}
In this section we extend the mirror conjecture to the case
of Calabi-Yau with bundles on top of it.  Let us be
physically concrete.   Consider a Calabi-Yau n-fold $W$ with
$N$ copies of $D_{2n}$ branes wrapped over it (here
we are considering Euclidean branes and so this is conventionally
denoted by $D_{2n-1}$ Euclidean brane--here it is more natural
to denote it by $D_{2n}$).  This gives rise to a $U(N)$ gauge bundle
on top of Calabi-Yau.  The condition to preserve supersymmetry
translates to having a stable bundle over $W$ which we will
henceforth assume.
Moreover the $c_1,c_2,...,c_{n}$ chern classes of the bundle
can be interpreted as the number (more precisely homology class) of
$D_{2n-2},D_{2n-4},...,D_0$
branes bound to the $D_{2n}$ brane \ref\doug{M. Douglas,
hep-th/9512077.}\ref\li{M. Li, Nucl. Phys. {\bf B460} (1996) 351.}.
The bundle should be non-degenerate if there is no common multiple
to $N,c_1,...,c_n$.
What should we expect the mirror to be?

Given that the mirror symmetry is at least morally
T-duality on each $T^n$ fiber, and T-duality
on D-branes exchanges Dirichlet and Neumann
boundary conditions the mirror of $N$ copies
of $D_{2n}$ brane will be an $n$-dimensional real submanifold
of $M$. Moreover if the bundle over $W$ is stable, as we assumed,
then the mirror should be a supersymmetric $n$ cycle (Lagrangian
relative to the Kahler 2-form and with minimal area).  Moreover
the homology class of the $n$-cycle is completely
fixed by $c_i$ of the bundle as there is a map from
each $(k,k)$ class of $M$ to an element of $H_n(W)$ (see
in particular \ref\oog{H. Ooguri, Y. Oz and Z. Yin,
Nucl. Phys. {\bf B477} (1996) 407.}\ which is very much
in the spirit of the present discussion).  In particular
the homology class of $C$, $H_n(C)$, is the mirror image of $\sum_i [c_i]$
(where we include
$c_0$ in this sum).

Let us assume that the bundle $V$ is non-degenerate over $W$ and so its
mirror is a smooth $n$-cycle $C\subset M$.  Apart from the above
relations between $V$ and $C$ there is one more important relation:
Since the resulting physics should be indistinguishable, this
implies that the moduli space of stable bundles $V$ with fixed $c_i$,
whose {\it complex} dimension is given by $H^1(End V)$ should be equal to the
dimension of moduli of the supersymmetric $n$-cycle $C$.  We now argue
that this in turn is equal to $H^1(C)$.  To see this note
that on the $D_n$ brane wrapped over $C$ we have a $U(1)$ bundle.
Thus the dimension of moduli of flat wilson lines is $b^1(C)$.
This, on the other hand should be paired with another moduli
to become the phase of a complex moduli
(similar to how the B-field and the real Kahler class combine).
We thus conclude that the complex moduli of the supersymmetric cylce
$C$, as far as its embedding in $M$ is concerned should be $H^1(C)$.
This is indeed the case, as the detailed analysis of \syz\
demonstrates.
Thus the complex dimension of the moduli space
of the supersymmetric cycle $C$ is $b^1(C)$
and we thus have an identification
$$H^1(End V,W)=H^1(C,M)$$
In fact more should be true, because the whole complex should
map to the corresponding mirror, i.e. we should expect for all $k$
$$H^k(End V,W)=H^k(C,M).$$

It is natural to choose ``special'' coordinates for the moduli
space of supersymmetric $C$ cycle as follows:  Let
$$\{ \gamma_j \} =H_1(C)$$
denote a basis for the 1-cycles of $C$.  Let $k$ denote
the Kahler class of $M$, which we decompose in terms
of an integral basis $\{ k_i \}$ of $H^2(M,{\bf Z})$
$$k=\sum_{i=1}^{h^{1,1}(M)} t_i k_i$$
Let us further assume that $H_1(M)=0$ (as is the case
other than for tori).  Then each $\gamma_j\in H_1(C)$
is contractible in $M$ and we choose a minimal area
disk $D_j\subset M$, which is holomorphic, whose boundary
is $\gamma_j$, i.e.
$$\partial D_j=\gamma_j$$
Define a real coordinate for the moduli of the supersymmetric cycle
$C$ by
$$r_j=\int_{D_j}k$$
i.e., the area of a minimal disk. Note that this will
not depend on the choice of the representative
$\gamma_j$, because if we
deform it to another 1-cycle $\gamma_j'$,  and consider
 $\Delta $ which is a 2 dimensional subspace
in $C$ with $\partial \Delta
=\gamma_j-\gamma_j'$, then the difference in ``area''
of the disks is $\int_{\Delta}k=0$, because $k$ restricted
to $C$ is zero (recall $C$ is Lagrangian relative to $k$).

Now, let the $U(1)$ holonomy
around $\gamma_j$ be denoted by $a_j$.  We combine
$r_j$ and $a_j$ to a ``special'' complex coordinate by
$${\tilde r}_j=r_j+i a_j$$
It is also natural to define the complex algebraic coordinates
\eqn\dco{{\tilde q}_j={\rm exp}(-{\tilde r}_j)}
Note that we now have as many complex coordinates for the
moduli of supersymmetric cycle $C$ as $b^1(C)$.  Note that
any other disk with boundary $\gamma_j$ will have an area
\eqn\are{A_j=r_j+\sum n_it_i}
with positive integers $n_i$.
This follows because the difference between two disks
bounding $\gamma_j$ is a closed  2-cycle in $W$ and so
$\int k_i$ over it gives the positive integer $n_i$.

Finally we apply this to Calabi-Yau threefold case and
write a formula generalizing \gnm .  In fact it is relatively
straight-forward to come up with the generalization.
The basic idea is that \gnm\ related computation of one
topological theory to another for the closed string sector
and we are after the open string sector of it.  The general
structure of the open string field theory in the target
space has been studied in \wcs.
On the complex side, $W$, we have a holomorphic version of
Chern-Simons theory  with action
\eqn\holcs{S=\int_W \Omega \wedge {\rm Tr} (A{\overline \partial}A +{2\over
3}A^3)}
where $A$ is a holomorphic (stable) $U(N)$  gauge connection which is
a $(0,1)$ form on Calabi-Yau threefold  $W$ with values in the
adjoint representation of $U(N)$.
On the $M$ side we have the usual
Chern-Simons theory corrected by sigma model instantons, i.e.,
\eqn\nhcs{S=\int_C{\rm Tr}({\cal A}d{\cal A}+{2\over 3}{\cal A}^3)+{\rm
Instanton}\ {\rm corrections}}
where the instanton corrections involve summing over holomorphic maps from
Riemann surfaces with boundaries, where boundary of the
Riemann surface is mapped to $C$, weighed by ${\rm exp}(-A)$
as well as with insertion of Wilson line on $C$. The specific
generalization of \gnm\ will correspond to computing
the disk worldsheet with three insertions of the $H^1$ fields of
$C$
at the boundary
(to rigidify the disk).
Setting these two contributions from the two
gauge theories equal will result in generalization
of \gnm .
The equality results in
\eqn\maneq{\int_W{\rm Tr}[ \delta_r A \delta_s A \delta_p A]\wedge
\Omega=\sum_{m_j,n_i}
d_{\{ m_j\} }^{\{ n_i\} }(r,s,p) \prod_{j=1}^{b^1(C)} {\tilde q}_j^{m_j}
\prod_{i=1}^{h^{1,1}(M)} q_i^{n_i}}
where $d_{\{ m_j\} }^{\{ n_i\} }(r,s,p)$ denotes the number of holomorphic
maps (up to a sign \wcs ) from a disk to $M$ where the boundary of the disk
is mapped to $C$ satisfying the following
conditions:  First, the class of this boundary curve
 in $H_1(C)$ is
$$\partial D=\sum_j m_j[\gamma_j].$$
Moreover, the class of $D-\sum  m_j D_j$, with $D_j$ defined
above, which is a
closed
2-cycle in $M$, is denoted by integers $n_i$ (relative to the integral
basis for $H_2(M)$).
Furthermore we require that
 three marked points on the boundary
of the disk get mapped in a definite cyclic order
to a point in specific 2-cycles $C^r,C^s,C^p\subset
C$ in that order,
which are Poincare dual to $\gamma_r,\gamma_s,\gamma_p$ in $C$.  
Note that just as in the usual mirror symmetry, the number
of such maps, in case there is a family of them, is to be
understood as the Euler class of some appropriate bundle
on moduli space.  Also one expects a similar modification
due to multi-wrapped disks, somewhat modifying the above formula
as in the usual case of mirror symmetry.  A special
case of the above relation is when $m_j=0=n_i$ in which
case on the Chern-Simons side one is getting
the classical triple intersection of the 2-cycles
$(r,s,p)$ on $C$.

The left-hand side of \maneq\ which involves derivatives
of the (0,1)-form connection $A$ relative to moduli of stable
bundles in directions r,s and p is a natural
generalization of variation of Hodge structure in the present
context.  Its structure in the above equation arises simply by
considering a three point function of the topological
open string B-model, and mapping it by mirror symmetry
to the corresponding three point function of the topological
open string A-model. 
Note that the
left-hand side of \maneq\ will depend on $q_i$ from the underlying
complex structure of Calabi-Yau and the fact that the
holomorphic connection $A$ knows about it,
as well as $\Omega$ in \maneq.  It also depends, of course
on the moduli of bundle which we are denoting by ${\tilde q}_j$.

Here we have only compared the contribution of
one open string worldsheet on the two sides.  One can of course
do this for arbitrary worldsheet geometries, just as in the
closed string case \bcov .  For example, just as in closed string
case \bcov\ we would
expect that if we considered the annulus instead of disk,
on the $W$ side we would be computing the Ray-Singer
torsion of the bundle $V$ over $W$ and on the $M$ side
the holomorphic maps from annulus to $M$ with boundary
being mapped to pairs of curves in $C$.  Clearly it would be
extremely interesting to work out some explicit examples in detail.

So far we have emphasized the role of the stable bundle
in solving an enumerative algebraic problem on the mirror.
One can also use this in the opposite direction.  Namely, instead
of thinking about bundles on the original manifold we might
as well study mid-dimension supersymmetric cycles on the mirror.
In the context of bundles on $K3$ this results in a surface
on the mirror and the study of this curve is equivalent
to studying the original bundle.  In this context this is well known
to mathematicians and in fact has been used in physical applications
\ref\fmw{R. Friedman, J. Morgan and E. Witten,
Comm. Math. Phys. {\bf 187} (1997) 679.}\ref\berset{M. Bershadsky,
A. Johansen, T. Pantev and V. Sadov, Nucl.Phys. {\bf B505} (1997) 165.}.
The generalization of this relation
between bundles and cycles on the mirror in higher
dimensions which we have found here is mathematically new.

We would like to thank Michael Bershadsky, Brian Greene, Maxim
Kontsevich and Eric Zaslow for
valuable discussions.
\vskip 1cm

This research is supported in part by NSF grant PHY-92-18167.

\listrefs
\end